\documentclass[showpacs,twocolumn,prb,superscriptaddress]{revtex4-1}
\usepackage{epstopdf}
\usepackage{color}
\usepackage{bm}
\usepackage{graphicx}
\usepackage{amsmath}
\usepackage{amsfonts}
\usepackage{amssymb}
\usepackage{bezier} 
\usepackage{color} 
\usepackage[colorlinks,bookmarks=false,citecolor=blue,linkcolor=red,urlcolor=blue]{hyperref}
\usepackage{times}

\graphicspath{{./Figs/}}

\begin{document}
\title{Numerical study of magnetization plateaux in the spin-1/2 Heisenberg antiferromagnet on the checkerboard lattice}
\author{Sylvain Capponi}
\affiliation{Laboratoire de Physique Th\'eorique, IRSAMC, Universit\'e de Toulouse, CNRS, F-31062, Toulouse, France}
\affiliation{Department of Physics, Boston University, 590 Commonwealth Avenue, Boston, Massachusetts 02215, USA}

\date{\today}
\begin{abstract}
We present numerical evidence that the spin-1/2 Heisenberg model on the two-dimensional checkerboard lattice exhibits several magnetization plateaux for $m=0$, $1/4$, $1/2$ and $3/4$, where $m$ is the magnetization normalized by its saturation value. These incompressible states correspond to somehow similar valence-bond crystal phases that break lattice symmetries, though they are different from the already established plaquette phase for $m=0$. Our results are based on Exact Diagonalization as well as Density Matrix Renormalization Group large-scale simulations, and interpreted in terms of simple parameter-free trial wavefunctions. 
\end{abstract} 

\maketitle

\section{Introduction}\label{intro}
Corner-sharing lattices, such as the kagom\'e or checkerboard ones in two-dimension (2d), or the hyperkagom\'e and pyrochlore lattices in three-dimension (3d), are known to be ideal playgrounds to study geometrical frustration. At the classical level, the Heisenberg model can we rewritten up to a constant as 
\begin{equation}
{\cal H}=J\sum_{\langle i j \rangle} {\boldsymbol{S}}_i \cdot {\boldsymbol{S}}_j =  \frac{J}{2} \sum_{\mathrm{simplex}\, \alpha} {\boldsymbol{S}}_\alpha^2 +\mathrm{Cst}
\end{equation}
where the second sum runs over all simplexes (triangles or tetrahedra) $\alpha$ and $J>0$ is the antiferromagnetic exchange. Thus, classical configurations must satisfy a local constraint $\boldsymbol{S}_\alpha=0,$ $\forall \alpha$, which implies a continuous degeneracy and an extensive entropy. This is the paradigmatic example of magnetic frustration leading to a disordered state for all temperatures.~\cite{Villain1979} Some famous examples are found in the three-dimensional pyrochlore lattice~\cite{Reimers1992,Moessner1998}, which can be realised in several materials.~\cite{Gardner2010} From now on, we will concentrate ourselves on the  two-dimensional (2d) analogue, known as the  checkerboard lattice (see Fig.~\ref{fig:vbc}), which is a square lattice of tetrahedra and thus more amenable to numerical simulations.

One expects on general grounds that quantum fluctuations will select some configurations among this manifold (so-called order by disorder phenomenon~\cite{Villain1980}). Indeed, spin-wave calculations have shown that the magnetically ordered N\'eel state is unstable towards a paramagnetic one for any spin $S$.~\cite{Canals2002} Early numerical studies on the $S=1/2$ Heisenberg model have pointed towards a non-magnetic state with a large spin gap~\cite{Palmer2001} (i.e. a plateau for zero magnetization), corresponding to some plaquette ordering~\cite{Fouet2003,Brenig2002,Tchernyshyov2003,Berg2003}. 
In principle, other (more exotic) phases are also possible when considering different couplings inter- and intra-tetrahedra, or anisotropic along the two axis~\cite{Tchernyshyov2003,Bernier2004,Starykh2005,Moukouri2008,Chan2011}.

In the presence of a magnetic field $h$, the Hamiltonian is simply changed to include a Zeeman term: 
\begin{equation}\label{eq:H+field}
{\cal H}=J\sum_{\langle i j \rangle} {\boldsymbol{S}}_i \cdot {\boldsymbol{S}}_j - h \sum_i S_i^z. 
\end{equation}
We will fix $J=1$ as the unit of energy in the following. 
For such a frustrated system, one generically expects magnetization plateaux to appear in the magnetization curve. This is simply because at the classical level for instance, the system remains very frustrated so that the situation is analogous to the zero-field case. 
For example, in the classical XY case, thermal fluctuations will select some  
uuud state, thus leading  to a plateau at $m=1/2$ (the magnetization $m=2\langle \sum_i S_i^z \rangle / N$, where $N$ is the number of sites, is normalized to its saturation value 1).~\cite{Canals2004} The same plateau can also be stabilized in a classical Heisenberg model on the 3d analogous pyrochlore lattice when spins are coupled to the lattice.~\cite{Penc2004} 
A similar mechanism can also occur in the quantum case: 
as an example, let us mention recent studies on the spin-1/2 Heisenberg model on the kagom\'e lattice where magnetization plateaux corresponding to incompressible phases that break lattice symmetries, so-called Valence Bond Crystal (VBC) states, have been established numerically for $m=1/3$, $5/9$ and $7/9$  in Refs.~\onlinecite{Capponi2013,Nishimoto2013}.

Another clue that plateaux may appear is provided via an \emph{exact localized magnon state} that can be constructed close to saturation field on such lattices.~\footnote{For a review, see Ref.~\onlinecite{Richter2004} and references therein.} 
 For the checkerboard lattice, one thus expects~\cite{Momoi2000,Oshikawa2000}  a plateau at $m=3/4$  corresponding to a 4-fold degenerate VBC with a finite gap to all excitations. 

Let us remind also that a commensurability criterion has to be satisfied in one-dimension~\cite{Oshikawa1997}, which was also generalised to any dimension~\cite{Hastings2004}: a unique featureless gapped state is possible iff $nS(1-m) \in \mathbb{Z}$ where $n$ is the number of sites in the unit cell, i.e. $n=2$ for the checkerboard lattice. Therefore, any plateau at finite $0 < m < 1$ must correspond either to a VBC with a larger unit cell $n>2$, or to a topological state. Clearly the latter possibility  is an exciting one, and indeed it was suggested for the $m=1/9$ plateau on the kagom\'e lattice.~\cite{Nishimoto2013} On the checkerboard lattice, our study will provide numerical evidence for several plateaux at $m=0$, $1/4$, $1/2$ and $3/4$, all of which corresponding to VBC phases that are either 2-fold or 4-fold degenerate depending on $m$. 

Let us mention that similar results were recently reported~\cite{Morita2016}  using density-matrix renormalization group (DMRG) computations, and will be discussed accordingly in Sec.~\ref{sec:dmrg}.

The outline of this paper is as follows: we will provide simple variational states that describe the various VBCs in Sec.~\ref{sec:ed} together with exact diagonalization results. Then, in Sec.~\ref{sec:dmrg}, we will present large-scale DMRG data as well as an extension to spin anisotropic interaction. Discussion and conclusion will be given in Sec.~\ref{sec:conclusion}.

\section{Variational states and Exact Diagonalization study}\label{sec:ed}
\subsection{Magnetization curves}\label{sec:mag}
We have performed extensive extensive diagonalization (ED) using Lanczos algorithm in order to
compute the magnetization curve for various lattices using periodic boundary conditions (PBC) to minimize finite-size effects. Basically, one simply needs to compute the total ground-state energy vs the total spin $S_z^{\mathrm{tot}}$ (without any magnetic field $h$) and then performs a Legendre transform to obtain $m(h)$. 

\begin{figure}
\includegraphics[width=0.6\columnwidth]{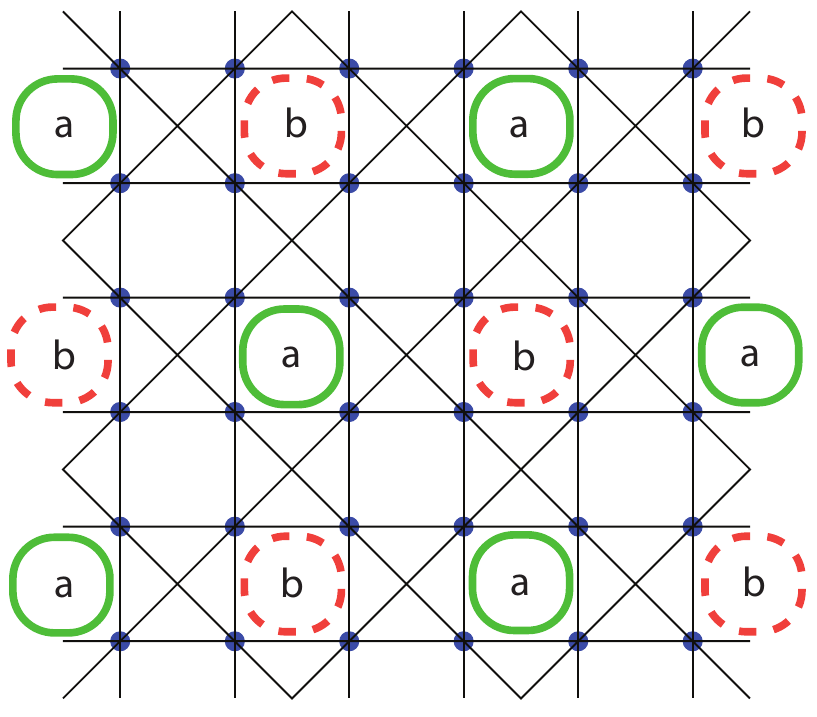}
\caption{(Color online) Sketch of the VBC phase where $a$ and $b$ denote eigenstates of a plaquette with total spin $a$ and $b$ respectively.  For instance, the product-state obtained with $(a,b)=(1,2)$ is an exact ground-state~\cite{Richter2004} for $m=3/4$. We will argue that similar phases with $(0,1)$ and $(0,2)$ are also realised for $m=1/4$ and $1/2$ respectively. In the $m=0$ case, the ground-state can be understood using $(0,0)$ product state~\cite{Fouet2003,Brenig2002,Berg2003}.}
\label{fig:vbc}
\end{figure}

As was mentioned already, exact localized ground-states can be built at $m=3/4$, and they correspond to the  pattern shown in Fig.~\ref{fig:vbc} (see Ref.~\onlinecite{Richter2004}), which possesses an 8-site unit cell. As a consequence, we have chosen clusters that can accomodate such VBC, such as $N=32$ (which has additional symmetry~\cite{Fouet2003}), $N=40$, and $N=64$. 
This localized magnon state corresponds to:
\begin{equation}\label{eq:psi34}
|\Psi_{\rm VBC}^{3/4}\rangle = \prod_j\vert L,\downarrow \rangle_j \prod_\ell\vert \uparrow \rangle_\ell,
\end{equation}
where the first product runs over an ordered pattern of all non-overlapping
 squares denoted by $a$ in Fig.~\ref{fig:vbc}
and the second product runs over the remaining $b$ sites. The localized-magnon state on a square is the ground-state with a single spin down.  Therefore, this exact VBC state can be viewed as a product state using $S_\square=1$ ground-state on plaquettes $a$ and $S_\square=2$ (i.e. polarized state) on $b$ ones. We will denote this product state as $(a,b)=(1,2)$.

Similarly to what we have observed on the kagom\'e lattice~\cite{Capponi2013}, we may construct similar product-state variational wavefunctions by simply flipping more spins on the $a$ or $b$ plaquettes, hence getting trial wavefunction at $m=1/2$ corresponding to $(0,2)$ or even $m=1/4$ for $(0,1)$. Of course, these are not exact eigenstates anymore, and nothing guarantees that they have any physical meaning for our microscopic model. So let us now present our unbiased numerical data. It is noteworthy that product states wavefunctions can naturally appear to describe some one-dimensional plateaux too.~\cite{Plat2015}

\begin{figure}
\includegraphics[width=\columnwidth]{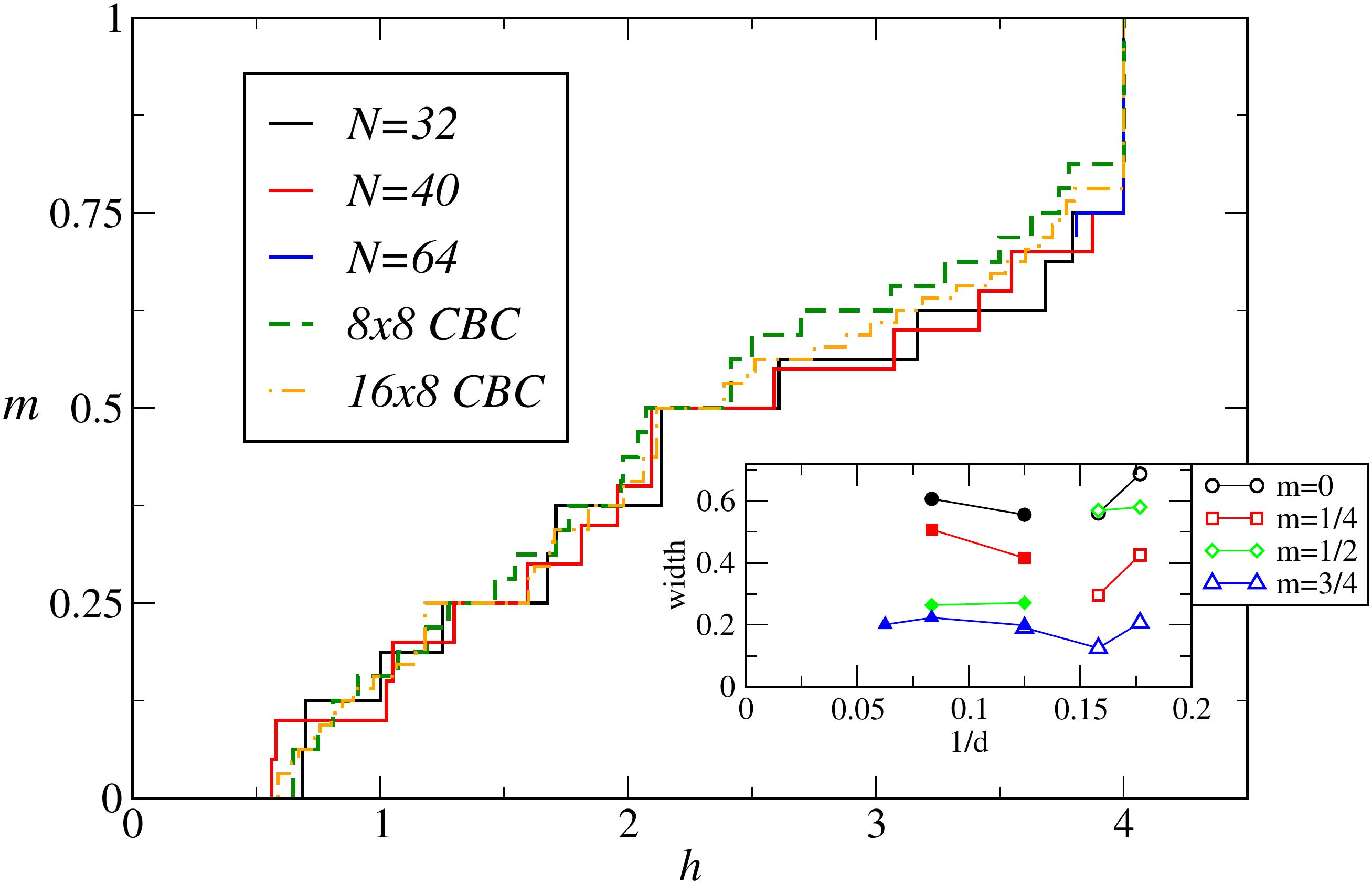}
\caption{(Color online) Magnetization curves of the $S=1/2$ Heisenberg model on the checkerboard lattice for clusters that can accommodate the VBC shown in Fig.~\ref{fig:vbc} for all the interesting $m$. Full/dashed lines correspond to ED with PBC / DMRG with CBC respectively, see text. Inset: Widths of the $m=0$, $1/4$, $1/2$ and $3/4$
  plateaux obtained from ED (open) and DMRG (full) on various clusters, plotted as a function of the inverse
  diameter. Data are consistent with finite values for all plateaux in the thermodynamic limit}
\label{fig:MofH}
\end{figure}

In Fig.~\ref{fig:MofH}, we plot the magnetization curve obtained on our particular choice  of clusters, thus extending data already published in Ref.~\onlinecite{Richter2004}. First, 
we recover some known features, such as the finite spin-gap ($m=0$ plateau) estimated to $\Delta \simeq 0.6 J$ previously~\cite{Fouet2003,Berg2003}, as well as the 
exact saturation field $h=4J$ that can be understood in terms of the
localized magnon eigenstates and a jump to $m=3/4$. Moreover, we observe several finite-size plateaux, including rather large ones for $m=1/2$ or $m=1/4$. A finite-size analysis for these values is performed in Fig.~\ref{fig:MofH} but it is rather difficult to conclude about the thermodynamic stability, especially for $m=1/4$. However, general arguments require that $m=3/4$ plateau should be finite~\cite{Momoi2000,Oshikawa2000} and we remind the reader that the spin gap (i.e. plateau for $m=0$) is believed to be finite from the literature, see above. 

\subsection{Energy spectroscopy from Exact Diagonalization}\label{sec:spectra}

Moreover, we can gain additional spectral signatures of these plateaux from ED data in the low-energy level excitations. More specifically, possible symmetry breaking can be investigated using the low-energy levels labelled by their quantum numbers. Since we are using periodic boundary conditions, we can label each eigenstate with magnetization $S^z$ and with their momentum, plus when allowed the irrep of the point-group symmetry. Note that we use the square lattice Brillouin zone, which has to be folded for the checkerboard lattice with 2-site unit cell.

\begin{figure}[!ht]
\includegraphics[width=0.95\columnwidth]{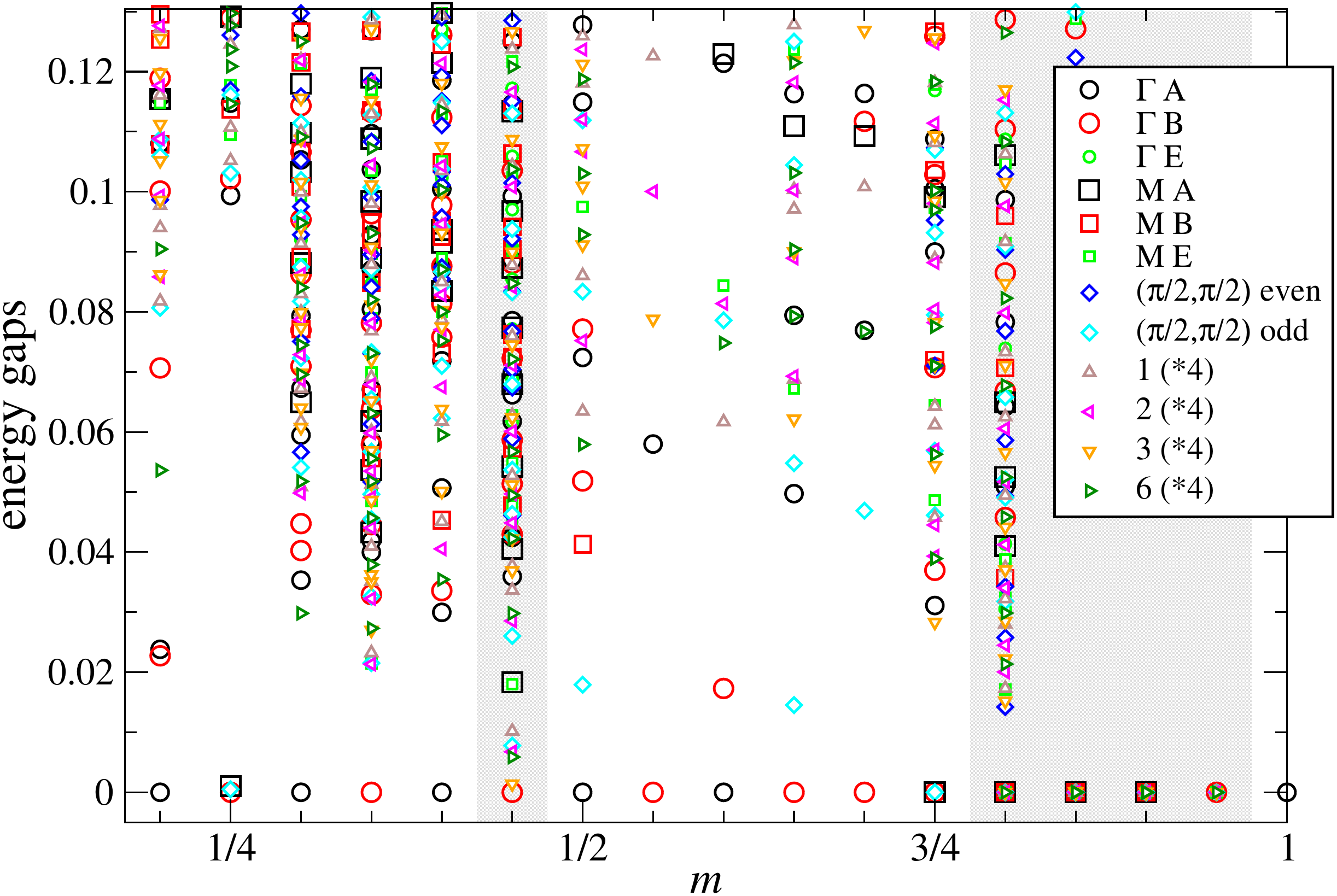}
\caption{(Color online) Energy gaps vs $m$ for $N=40$ lattice labeled with
their quantum numbers. Relevant momenta are labelled as $\Gamma=(0,0)$, $M=(\pi,0)$ and 2-fold degenerate $(\pi/2,\pi/2)$. For 
$m=1/4$, $1/2$ and $3/4$, the lowest states correspond to the expected four ones  in the Brillouin zone. 
The magnetization sectors with a gray background are not visited in the magnetization curve, i.e., they are obscured by a magnetization jump.}
\label{fig:gaps}
\end{figure}

For the proposed 4-fold degenerate VBC states at finite $m$, a symmetry analysis leads to 4 degenerate states: one  at the $\Gamma$ point, one at $M=(\pi,0)$ and one at each (two-fold degenerate) $(\pi/2,\pi/2)$ momentum. 
 In Fig.~\ref{fig:gaps}, we plot the energy gaps for each magnetization sector on $N=40$ cluster, and we do observe a very good (quasi) degeneracy of these 4 states and a sizeable gap above for $m=1/4$ and $m=3/4$. At $m=1/2$, these are also the lowest states but the separation is less clear, presumably due to a larger correlation length.

\subsection{Correlations}\label{sec:correlations}

Another piece of evidence comes from the computation of specific correlations using the unique finite-size ground-state. More specifically, 
for each magnetization $m$, we have computed connected spin correlation functions 
\begin{equation}\label{eq:szsz}
\langle S_i^z S_j^z\rangle_c = \langle S_i^z S_j^z\rangle - \langle
S_i^z\rangle \langle S_j^z\rangle
\end{equation}
as well as connected dimer-dimer correlations
\begin{eqnarray}\label{eq:dimerdimer}
&&\langle (\boldsymbol{S}_i\cdot \boldsymbol{S}_j) (\boldsymbol{S}_k\cdot \boldsymbol{S}_\ell) \rangle_c  \nonumber\\
&&= \langle (\boldsymbol{S}_i\cdot \boldsymbol{S}_j) (\boldsymbol{S}_k\cdot \boldsymbol{S}_\ell) \rangle  - \langle (\boldsymbol{S}_i\cdot \boldsymbol{S}_j)\rangle \langle (\boldsymbol{S}_k\cdot \boldsymbol{S}_\ell) \rangle
\end{eqnarray}
using ED on various lattices. 

\begin{figure*}
\includegraphics[width=0.235\linewidth]{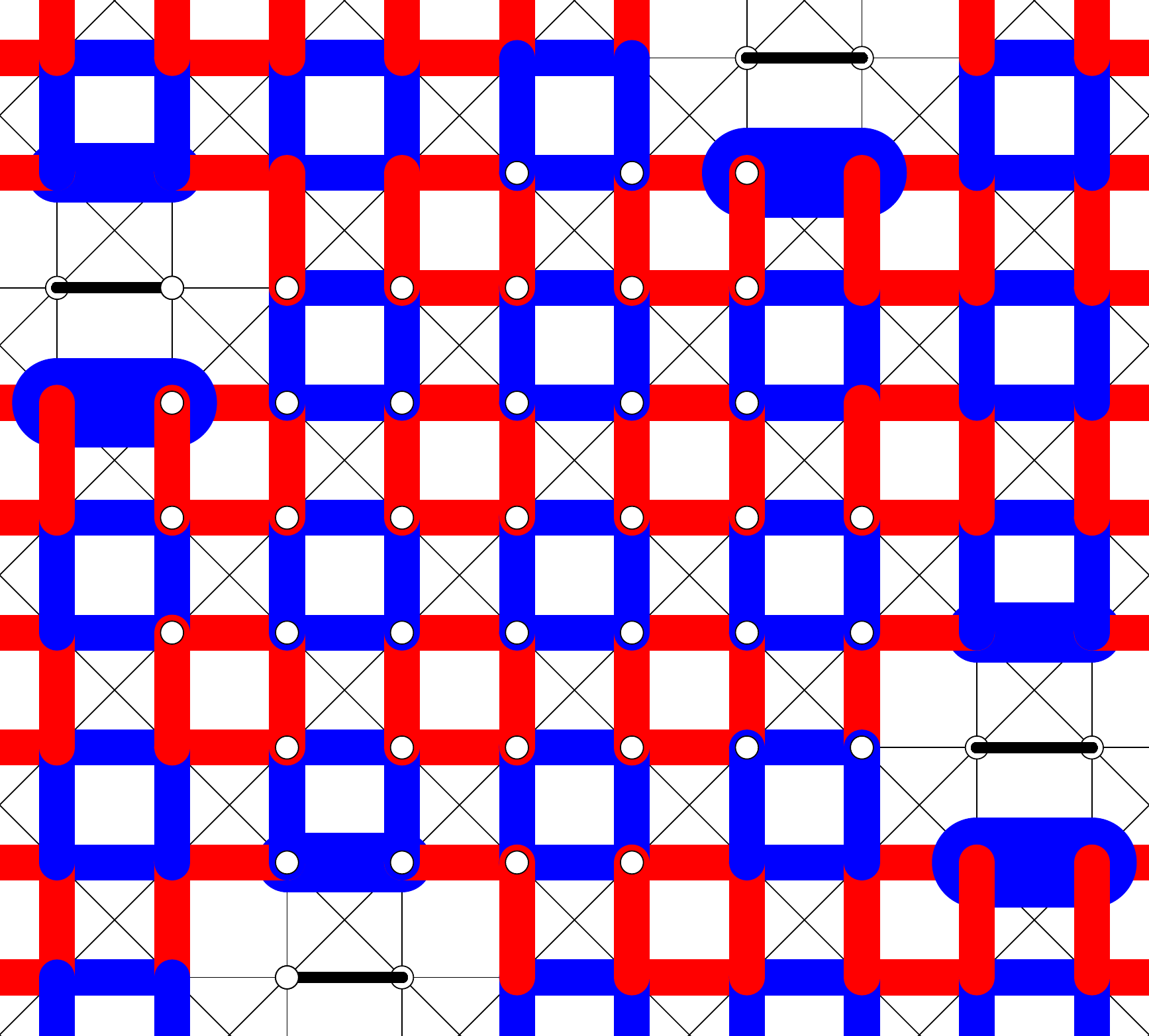}
\hspace{0.2cm}
\includegraphics[width=0.235\linewidth]{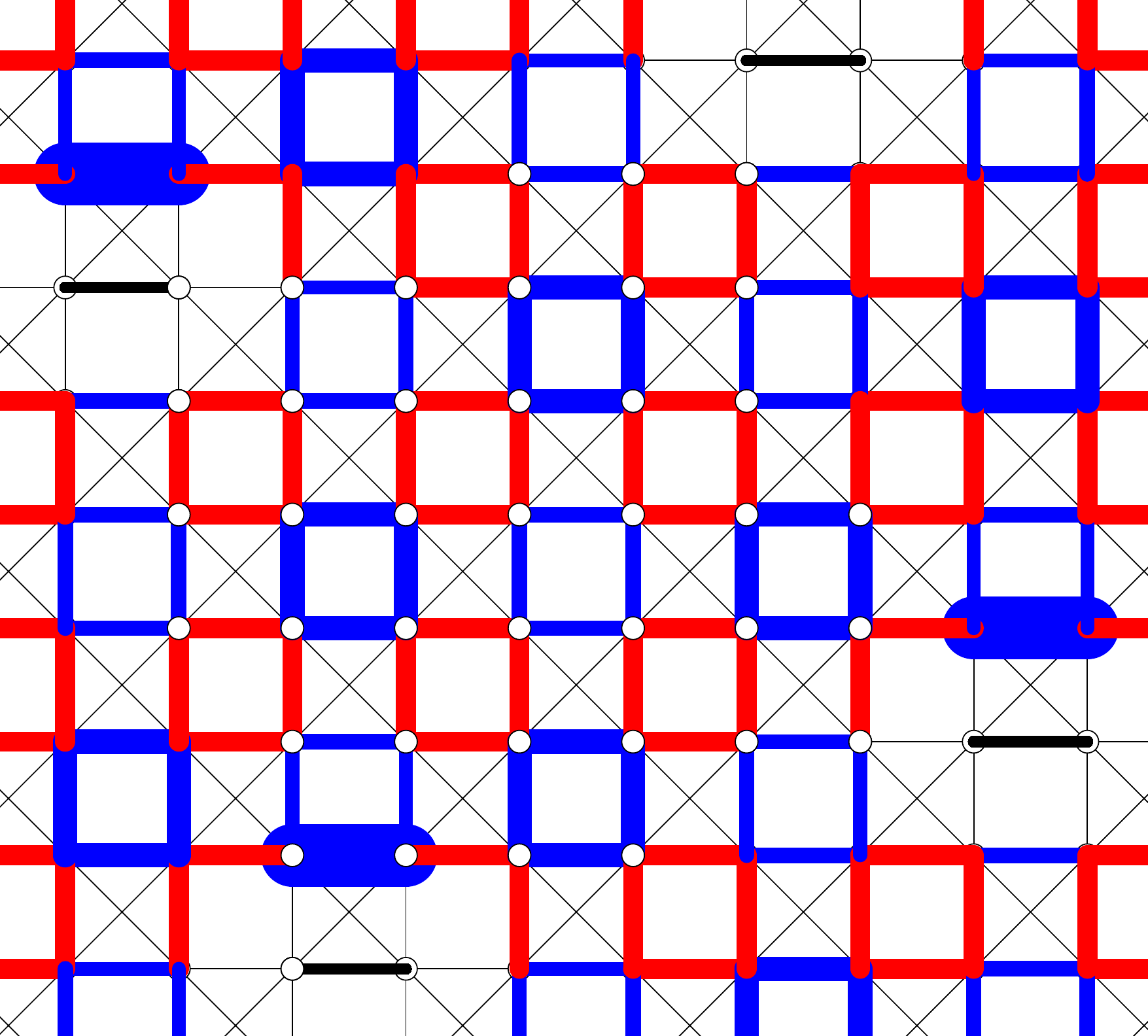}
\hspace{0.2cm}
\includegraphics[width=0.235\linewidth]{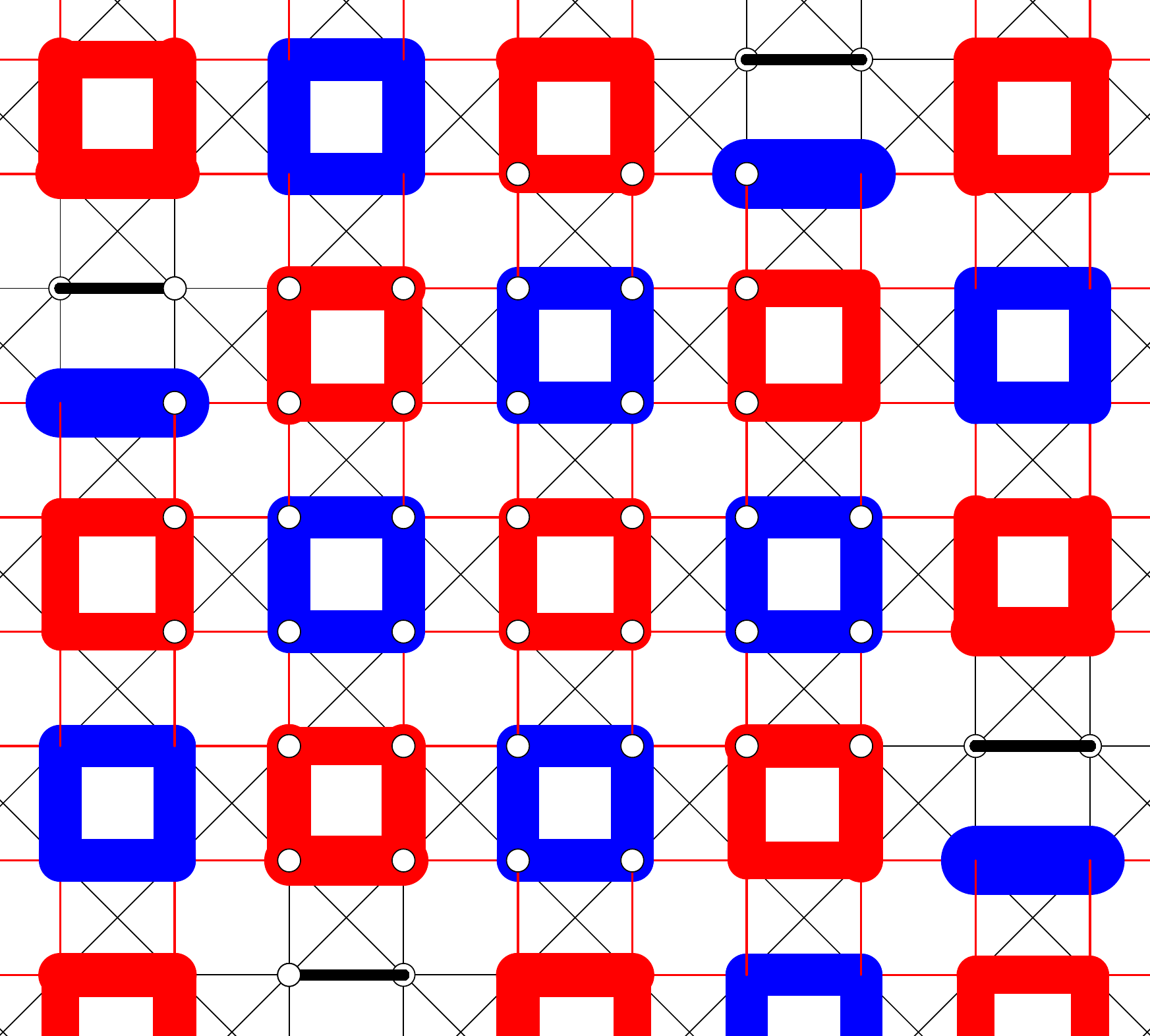}
\hspace{0.2cm}
\includegraphics[width=0.235\linewidth]{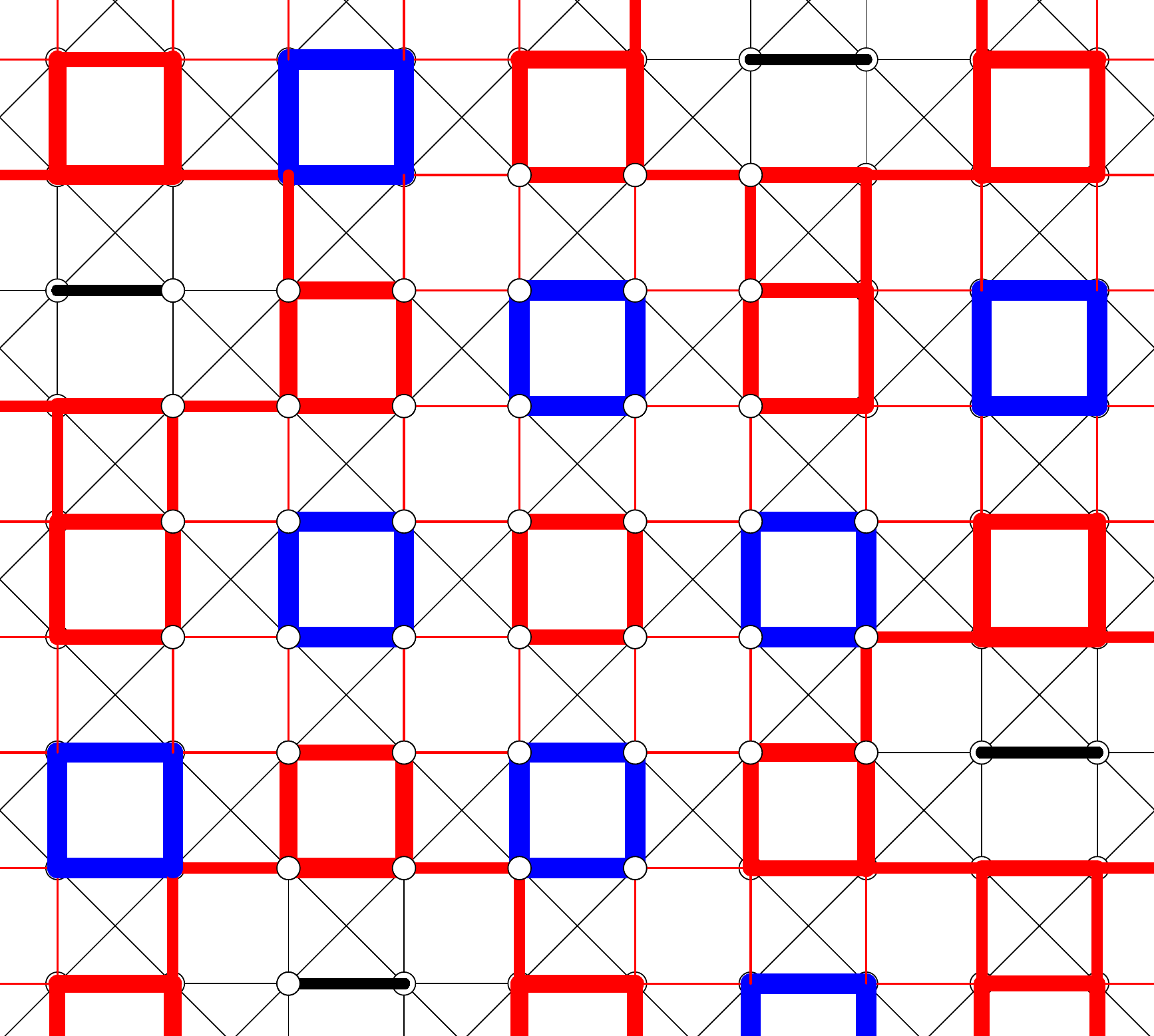}

\vspace{0.4cm}

\includegraphics[width=0.235\linewidth]{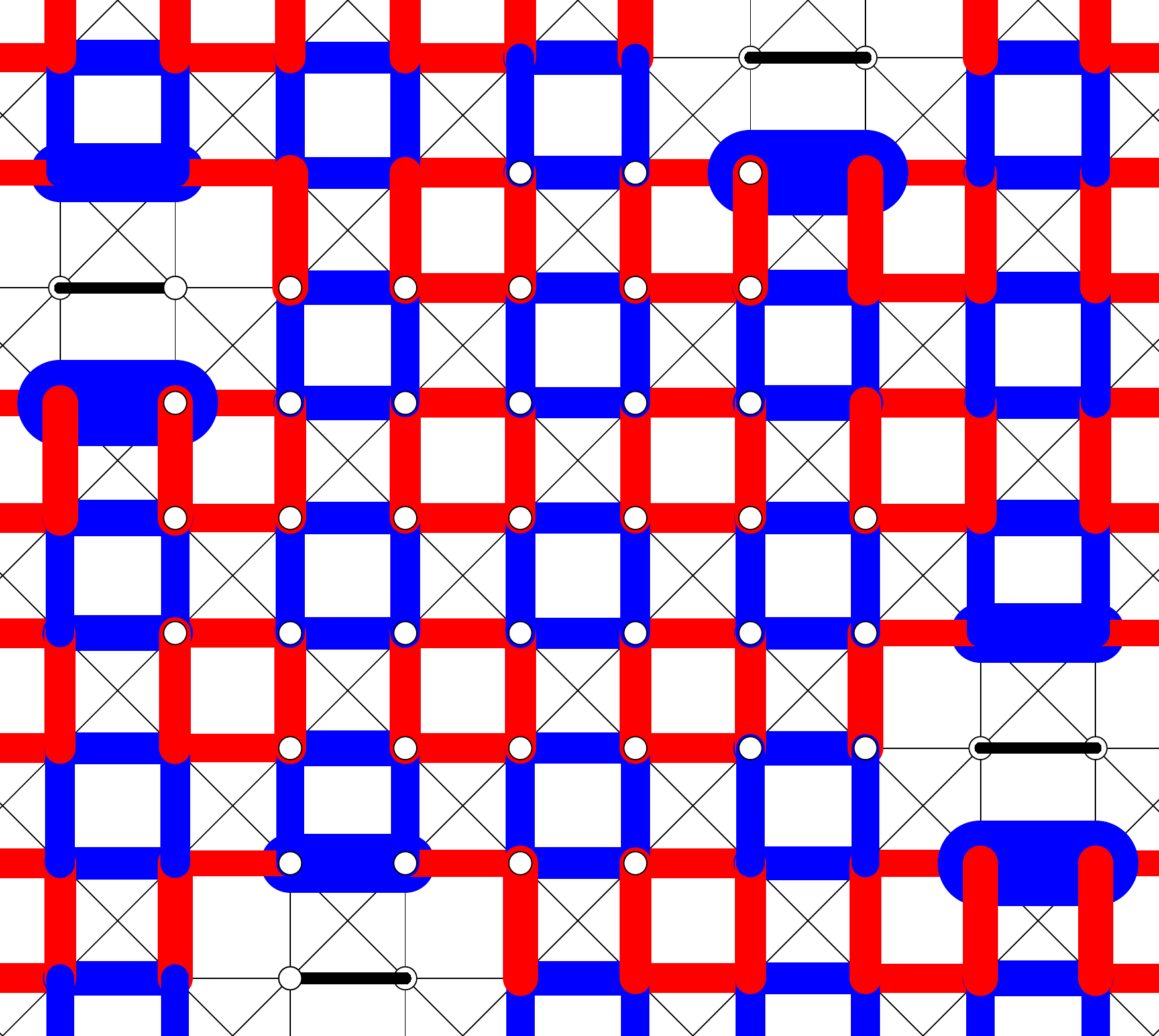}
\hspace{0.2cm}
\includegraphics[width=0.235\linewidth]{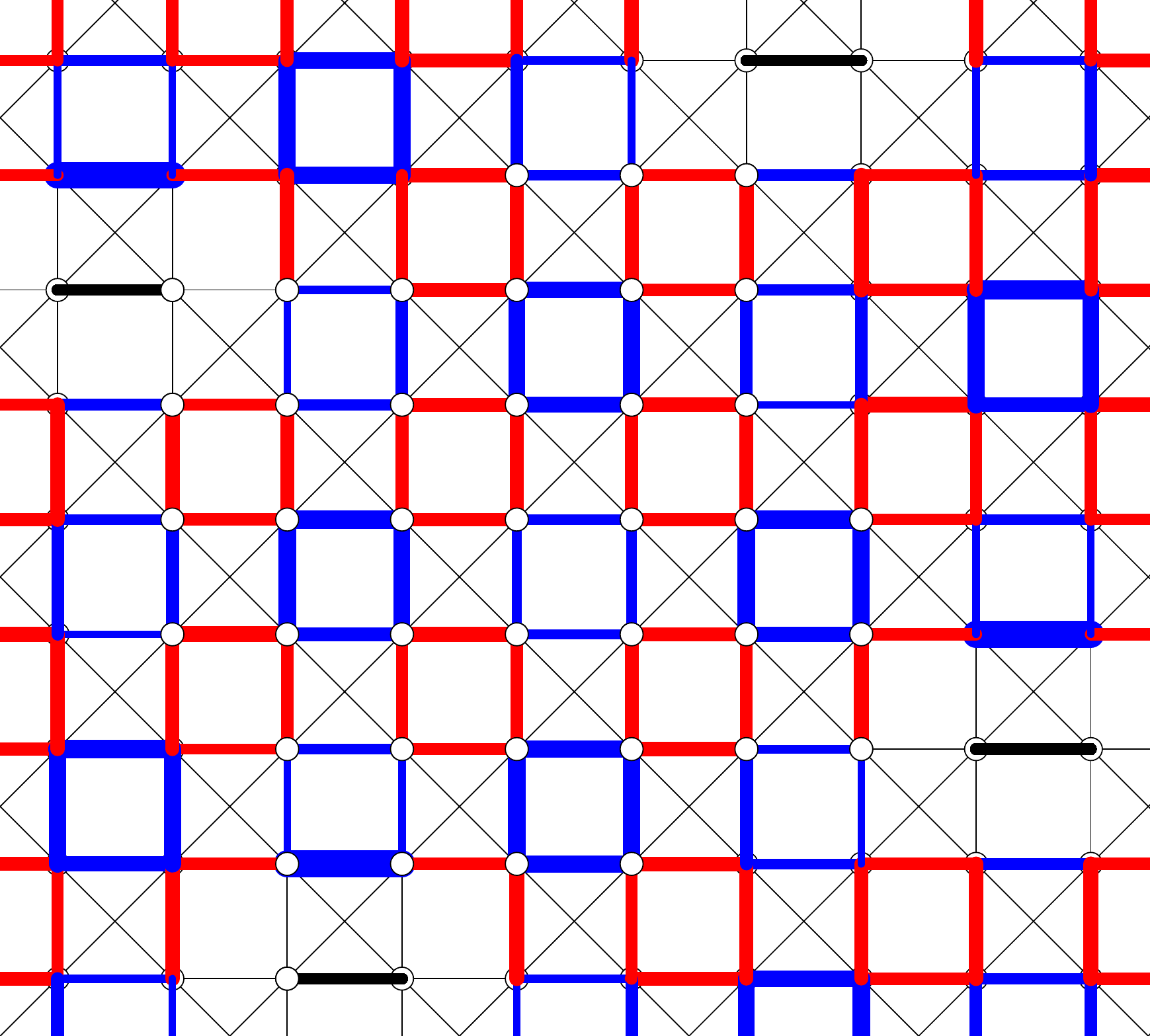}
\hspace{0.2cm}
\includegraphics[width=0.235\linewidth]{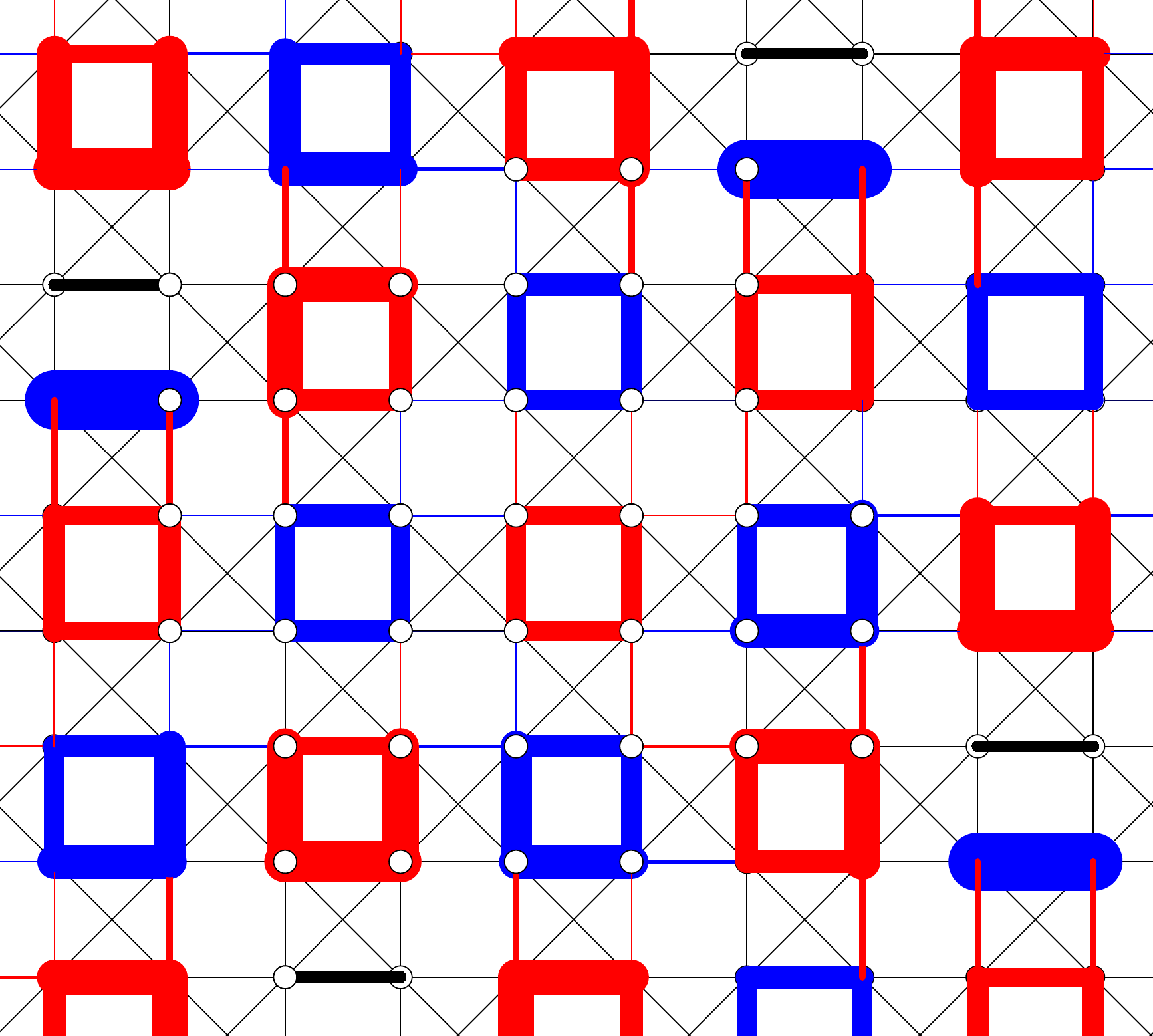}
\hspace{0.2cm}
\includegraphics[width=0.235\linewidth]{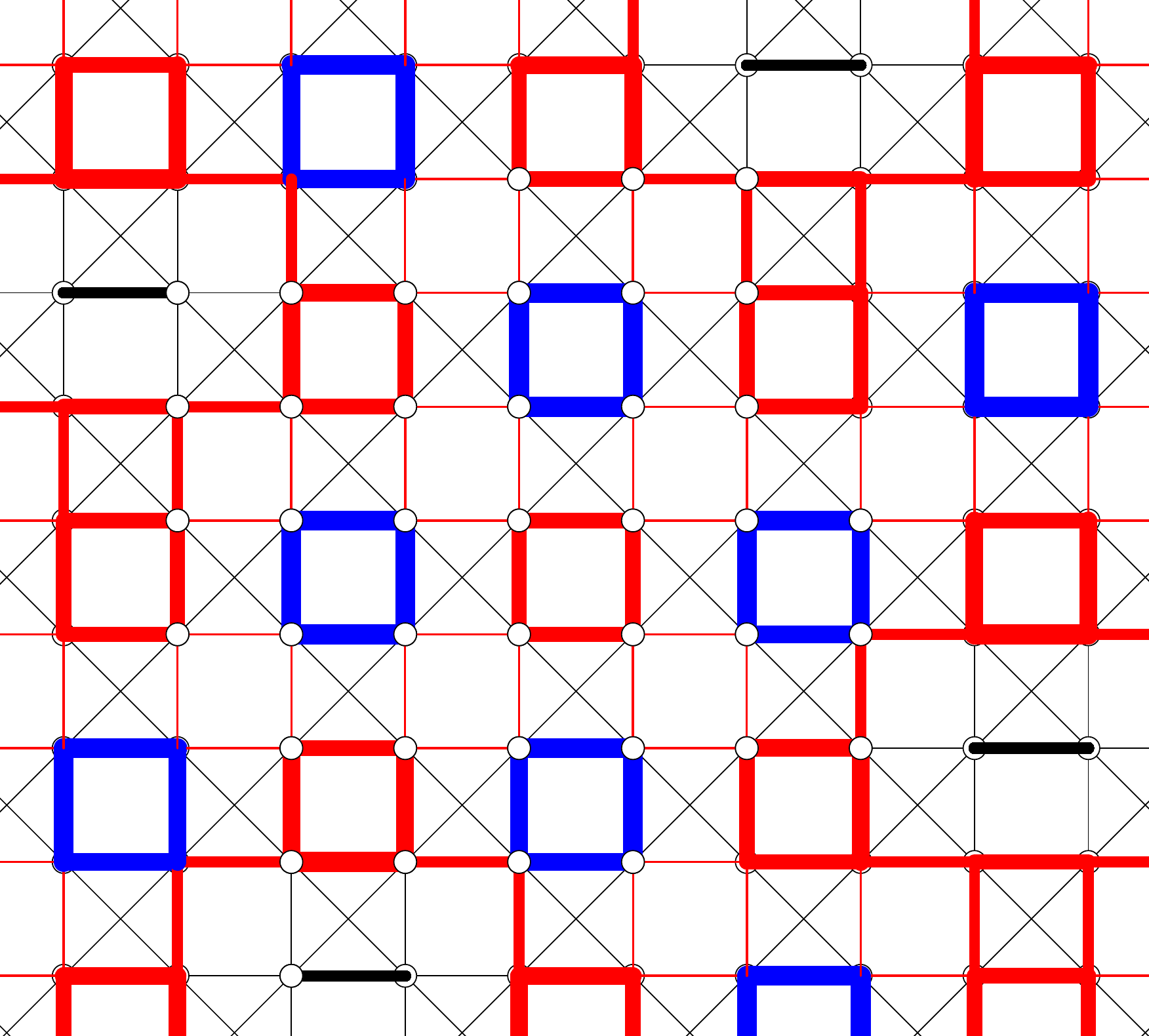}
\caption{(Color online) Dimer-dimer correlations (cf.~Eq.~(\ref{eq:dimerdimer})) computed on $N=40$ cluster: positive and negative values are shown respectively with   blue or red lines,  and width is proportional to the data; reference bond is
 shown in black. From left to right, data correspond to the ground-state for $m=0$, $1/4$, $1/2$ and $3/4$. 
 The top row corresponds to simple VBC states where computation is done analytically without any free parameter (see text); bottom row corresponds to ED results for the Heisenberg model. 
 }
\label{fig:CorrPanel}
\end{figure*}

Similarly to what was done on the kagome lattice~\cite{Capponi2013}, one can obtain exact expression on the $m=3/4$ VBC state, see Fig.~\ref{fig:vbc} or Eq.~\eqref{eq:psi34}. Discarding short-distance data that require a separate calculation, one can easily get that: (i) Average $\langle \boldsymbol{S}_i \cdot \boldsymbol{S}_j \rangle = 1/16$; (ii) connected correlations between ``strong'' plaquette bonds are $9/256 \simeq 0.035$; (iii) connected correlation between strong and weak plaquette bonds are $-7/256 \simeq -0.027$. These are exactly the numbers we get in Fig.~\ref{fig:CorrPanel} (with small deviations due to a finite overlap of the four localized magnon states on a finite cluster). A similar calculation can be performed for any trial state $(a,b)$ and in Fig.~\ref{fig:CorrPanel}, we compare ED data to simple VBC, respectively given as: $(a,b)=(0,0)$ for $m=0$, $(0,1)$ for $m=1/4$, $(0,2)$ for $m=1/2$, and $(1,2)$ for $m=3/4$. In all cases, we do observe a (semi-)quantitative agreement, which strongly supports the description of all these plateaux states in terms of simple VBC as sketched in Fig.~\ref{fig:vbc}.

\section{DMRG results}\label{sec:dmrg}

\subsection{Simulations on large cylinders}
 We now turn to large-scale simulations using 2d Density Matrix Renormalization Group algorithm (DMRG)~\cite{White1992} which can be applied on cylinders provided that the width is not too large. We have kept up to  $m=6000$ states in order to obtain good convergence of our results and a small discarded weight. We use cylindrical boundary conditions (CBC). 

In Fig.~\ref{fig:MofH}, we have plotted the magnetization curve on few cylinders which exhibit large plateaux for the expected magnetizations. Note that the $m=3/4$ plateau is slightly shifted due to boundary effects, so that we have systematically taken the plateau next to the jump to saturation instead (which converged to $3/4$ in the thermodynamic limit). A finite-size scaling can be performed to check that all proposed 4 VBCs are stable in the thermodynamic limit, see inset of Fig.~\ref{fig:MofH} with data obtained on $12\times 8$, $16\times 12$ and $16\times 16$ cylinders.

\begin{figure*}
\includegraphics[width=0.45\linewidth]{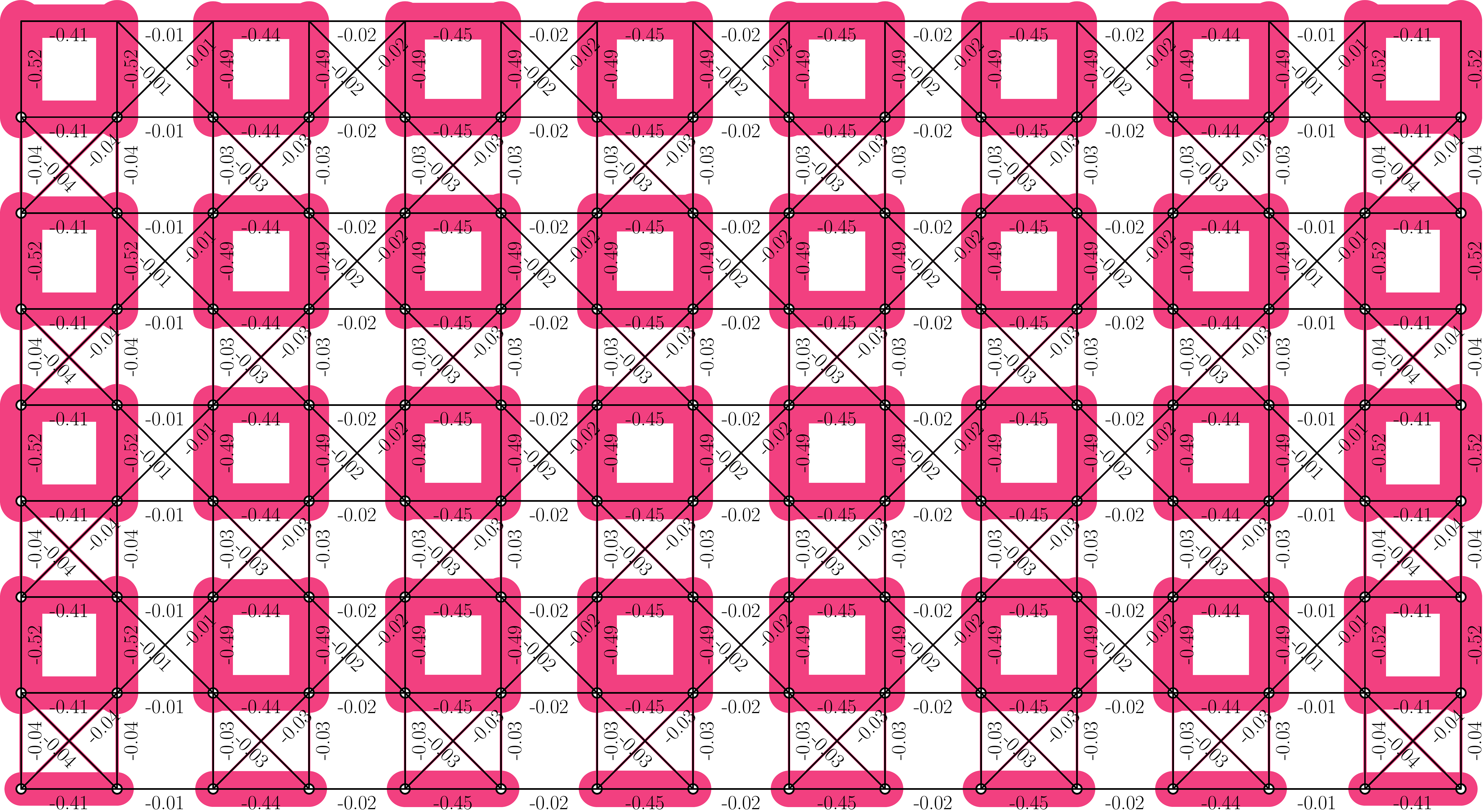}
\hspace{0.2cm}
\includegraphics[width=0.45\linewidth]{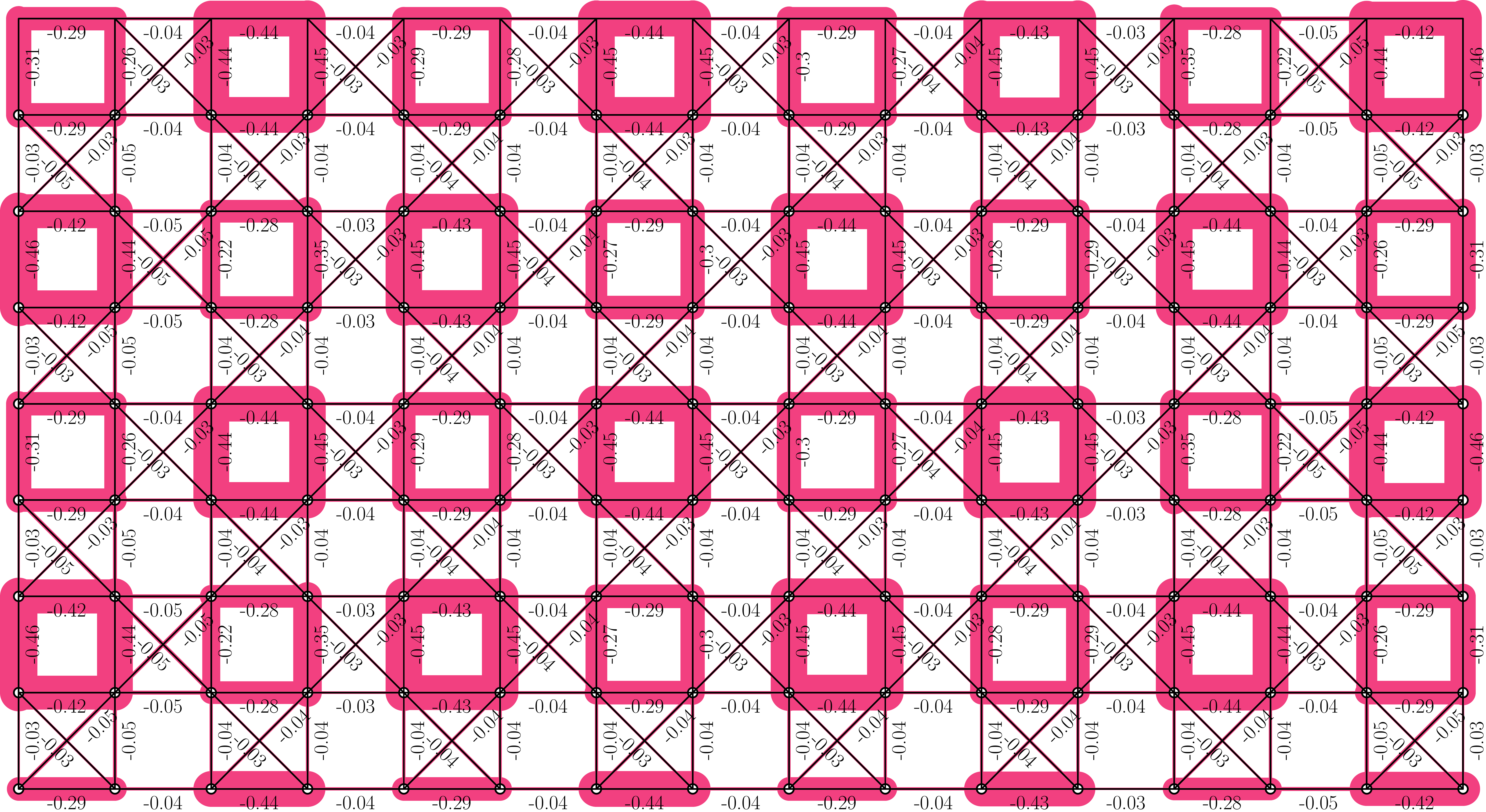}
\includegraphics[width=0.45\linewidth]{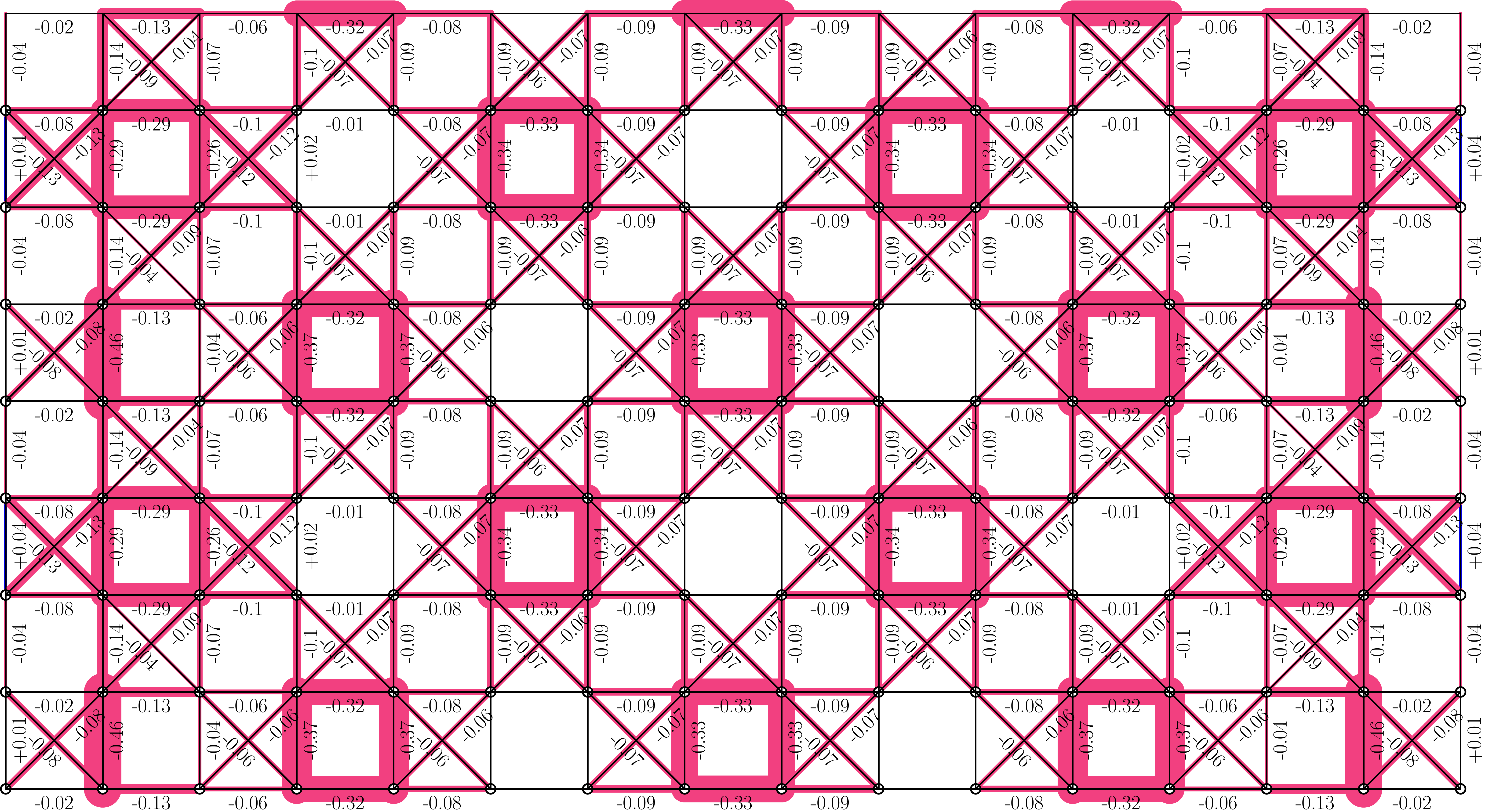}
\hspace{0.2cm}
\includegraphics[width=0.45\linewidth]{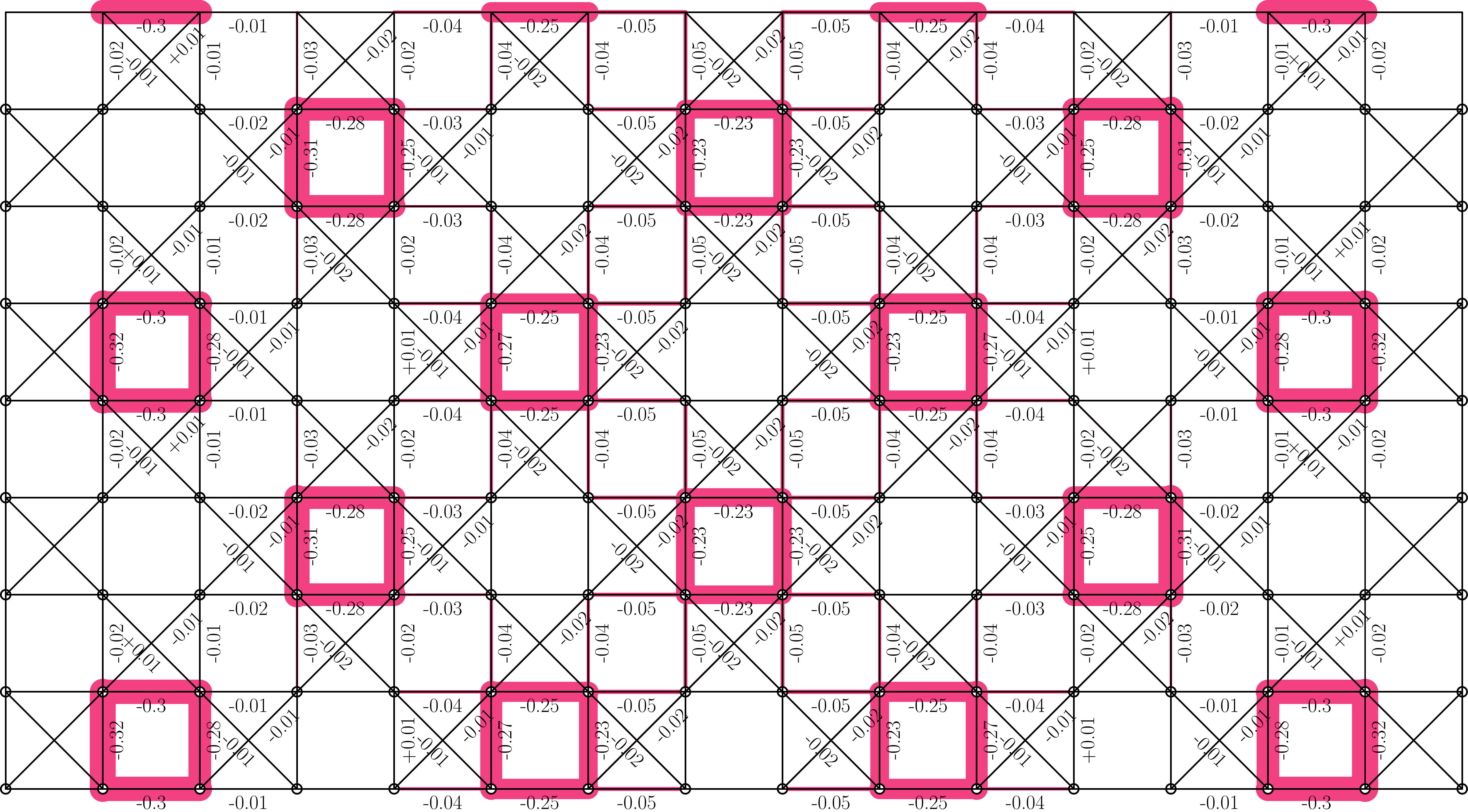}
\caption{(Color online) Bond strengths for $m=0$, $m=1/4$, $m=1/2$ and $m=3/4$ (from top to bottom and left to right) on $16\times 8$ cylinder using DMRG simulations and keeping up to 6,000 states. In each plot, top and bottom lines are identical due to periodic boundary conditions along this direction (CBC). Data larger than $0.01$ (in absolute value) 
are written on the plot.   
}
\label{fig:dmrg1}
\end{figure*}

\begin{figure}[!h]
 \includegraphics[width=\linewidth]{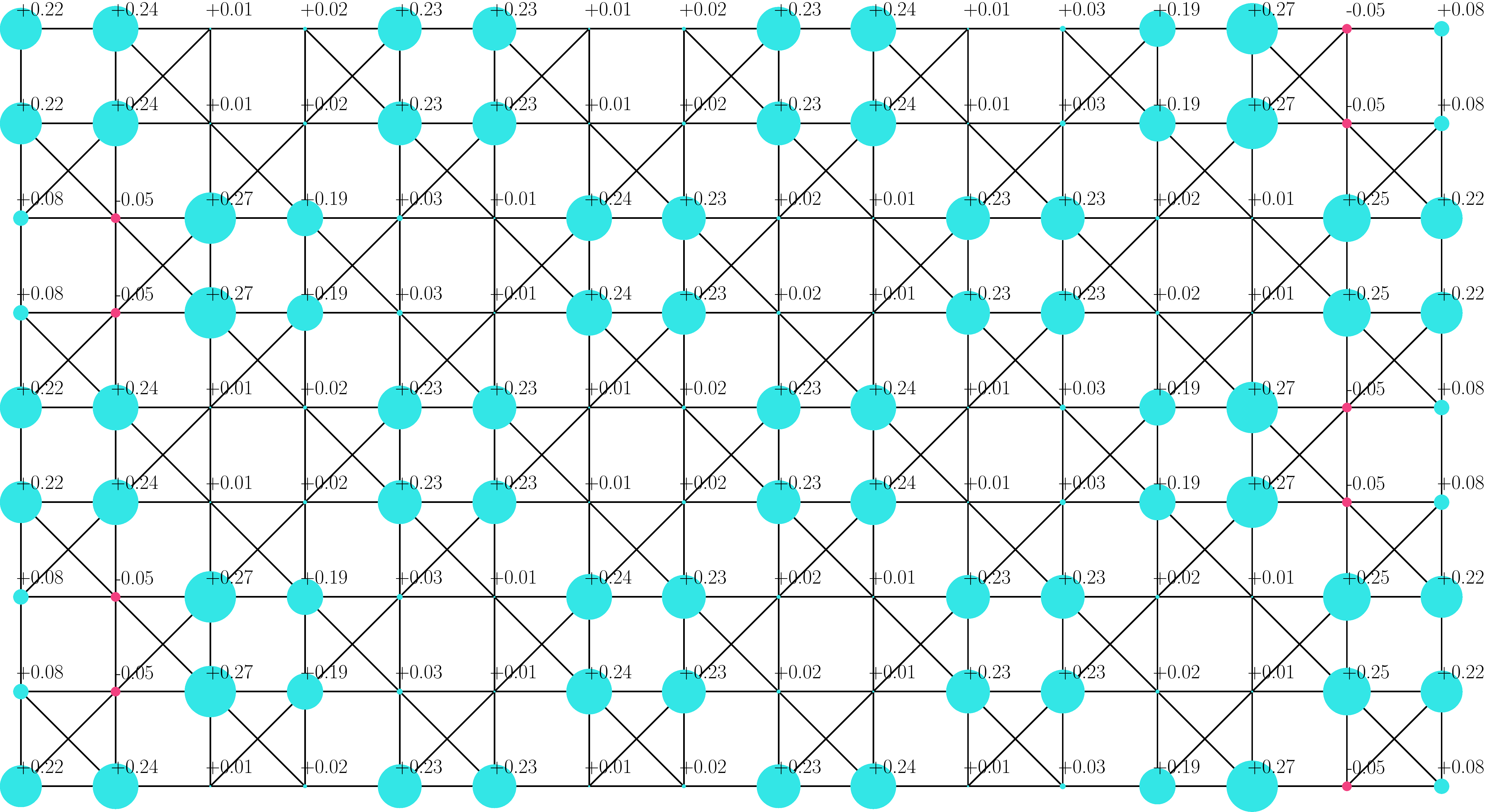}
\includegraphics[width=\linewidth]{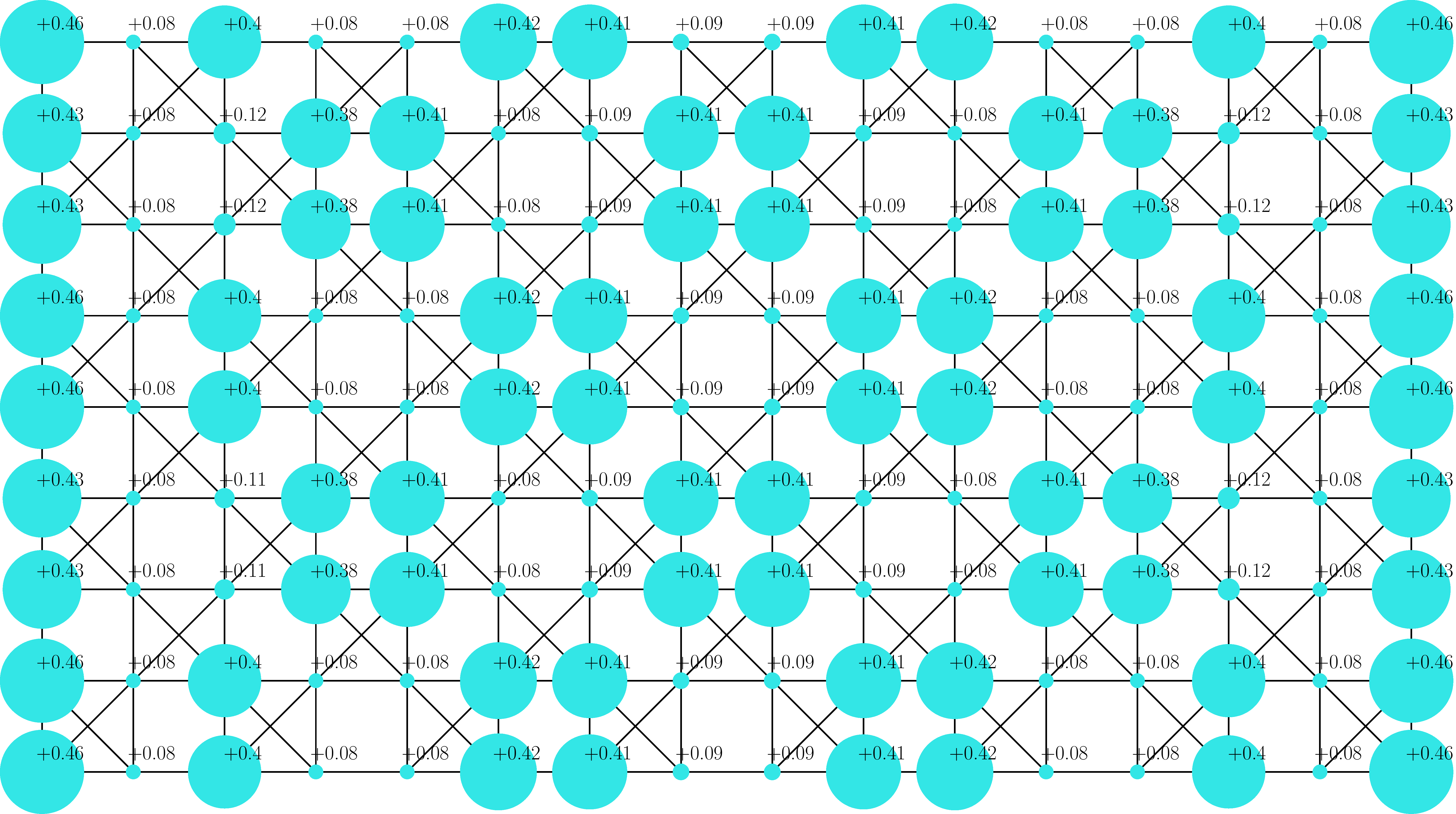}
\includegraphics[width=\linewidth]{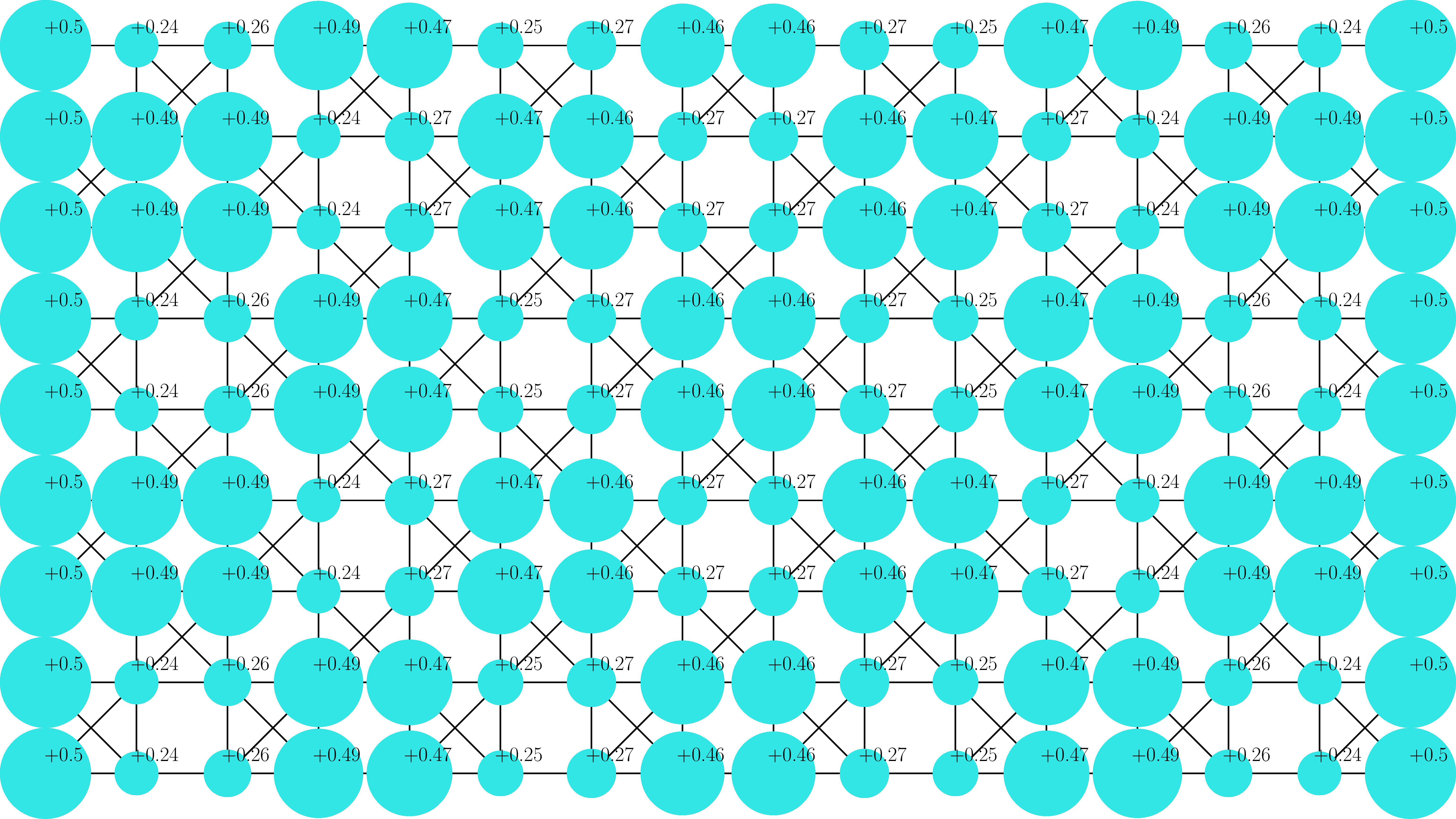}
\caption{(Color online) Local magnetization for $m=1/4$, $m=1/2$ and $m=3/4$ (from top to bottom) on $16\times 8$ cylinder using DMRG simulations and keeping up to 6000 states. Radius is proportional to the absolute value. Due to edge effects, for $m=3/4$ plot, we have used a total magnetization slightly higher ($m=3/4+2/128$) to get a clearer picture of the bulk behaviour. Top and bottom lines are identical (CBC). }
\label{fig:dmrg2}
\end{figure}

In Fig.~\ref{fig:dmrg1}, we plot the local bond strengths  
$\langle {\boldsymbol{S}}_i \cdot {\boldsymbol{S}}_j\rangle - \langle S^z_i\rangle  \langle S^z_j\rangle $ for various magnetizations. 
Data are compatible with a 4-fold degenerate ground-state at $m=1/4$, $m=1/2$ and $m=3/4$, but only 2-fold for $m=0$, as expected from our trial wavefunction guess. We remind the reader that the DMRG algorithm will target one particular VBC, and not a superposition, since it converges to so-called minimally entangled states.~\cite{Jiang2012}  In addition, while for $m=0$, all sites have a vanishing magnetization on average ($\langle S_i^z\rangle=0 \, \forall i$), which is an exact result~\cite{Lieb1999}, we have found a clear modulation in this quantity for all $m>0$ plateaux states in agreement with the proposed VBC states. More precisely, 
for the plateaux corresponding to $m=1/4$, $1/2$ and $3/4$, we have argued that they could be understood using simple product states with total spin $a$ and $b$ on alternating 4-site plaquettes. Respectively, we have proposed $(a,b)=(0,1)$, $(0,2)$ and $(1,2)$, see Fig.~\ref{fig:vbc}. 
So in principle, using a DMRG simulation on cylinders that will target one particular state, \emph{we can directly measure the local magnetization} $\langle S_i^z \rangle$ to characterize the state. This is shown in Fig.~\ref{fig:dmrg2} for a $16\times 8$ cylinder. In all cases, the total magnetization on each 4-site plaquette is very close to the expected one in a simple product state, i.e. $S_{\square}^z=0$, 1 or 2. We note also that there are strong edge effects so that one should be cautious with boundary conditions and extrapolations. This is also why for instance, we have fixed $m=3/4+2/N$, where $N$ is the number of sites, so that the bulk resembles more the genuine $m=3/4$ ground-state.

Similar results for the bond pattern and the local magnetizations have been obtained in Ref.~\onlinecite{Morita2016} using DMRG as well but with modified boundary conditions to minimize finite-size effects, at the cost of having to fix the magnetic field and not the total magnetization (grand-canonical approach). Both studies agree on the existence and nature of the plateaux phases at m=0, 1/4, 1/2 and 3/4. Moreover, Morita and Shibata argue in favor of a small plateau for m=3/8, and maybe also at m=1/8 where they observe a small anomaly.

\subsection{Quantum phase transition away from the SU(2) case}\label{sec:qpt}

In order to make connection with related strongly correlated models on the same lattice, it turns out to be useful to investigate the (spin) anisotropic XXZ model: 
\begin{equation}
\label{eq:XXZ}
{\cal H} = \sum_{\mathrm{bonds}\, (i,j)} \frac{1}{2}\left(S_i^+ S_j^- + S_i^- S_j^+\right) + \Delta S_i^z S_j^z,
\end{equation}
where $\Delta$ quantifies the anisotropy.  For this part, we will focus  on the $m=1/2$ plateau and we will show how increasing $\Delta$ leads to a quantum phase transition and a qualitative change in the physical properties. 

Indeed, for strong $\Delta\gg 1$, the models maps onto a purely kinetic quantum dimer model (QDM), which is defined on the new square lattice which sites are on the center of the crossed plaquettes of the original one. The mapping simply consists in associating a dimer for each $\downarrow$ spin and the hardcore constrains simply reflects the ice-rule (one down spin per crossed plaquette). The QDM ground-state is known to be columnar-like~\cite{Syljuasen2006,Banerjee2014,Schwandt,Banerjee2016}, so that the ground-state is also 4-fold degenerate but of a different kind.

\begin{figure}[!h]
\includegraphics[width=\linewidth]{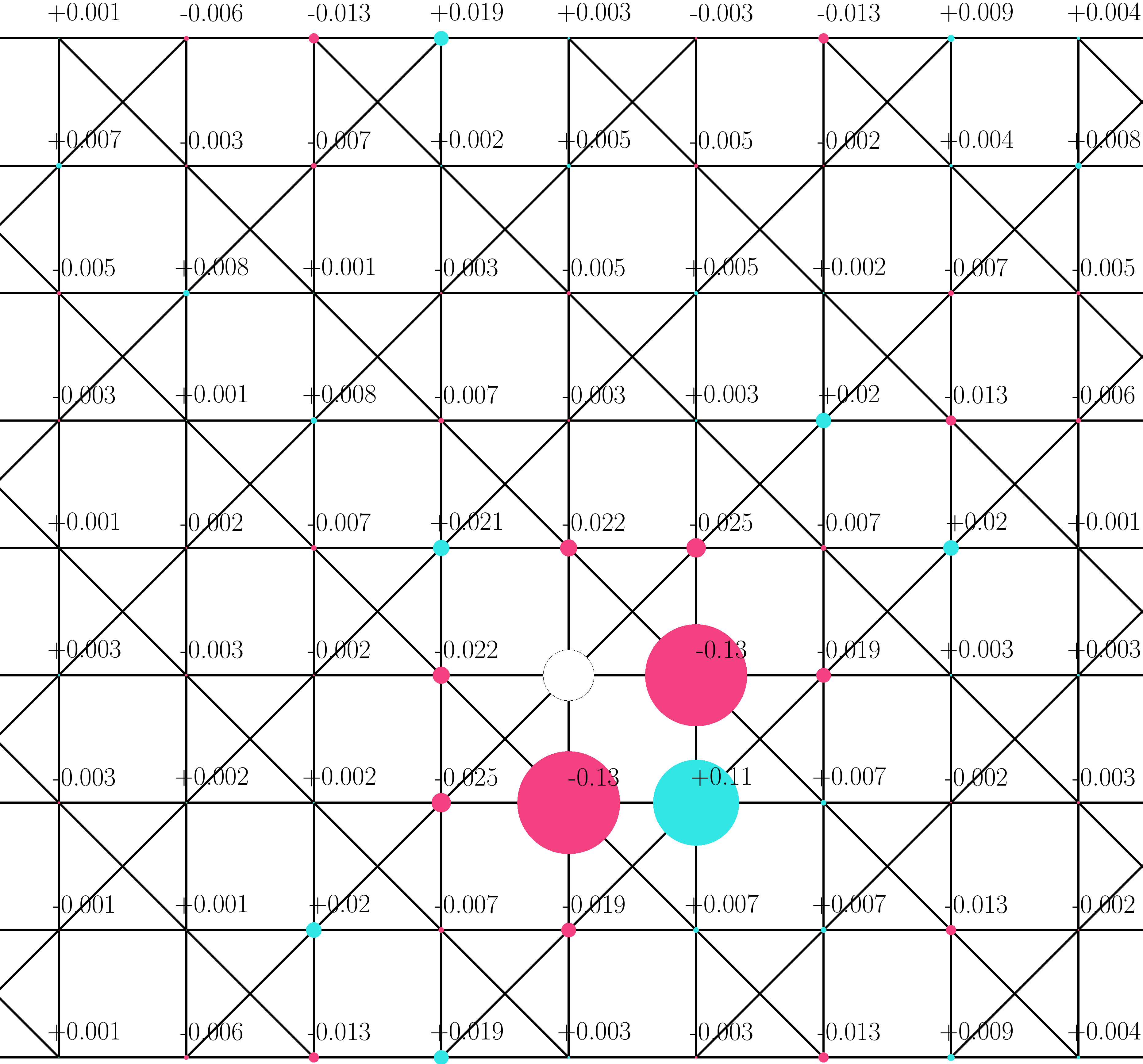}
\includegraphics[width=\linewidth]{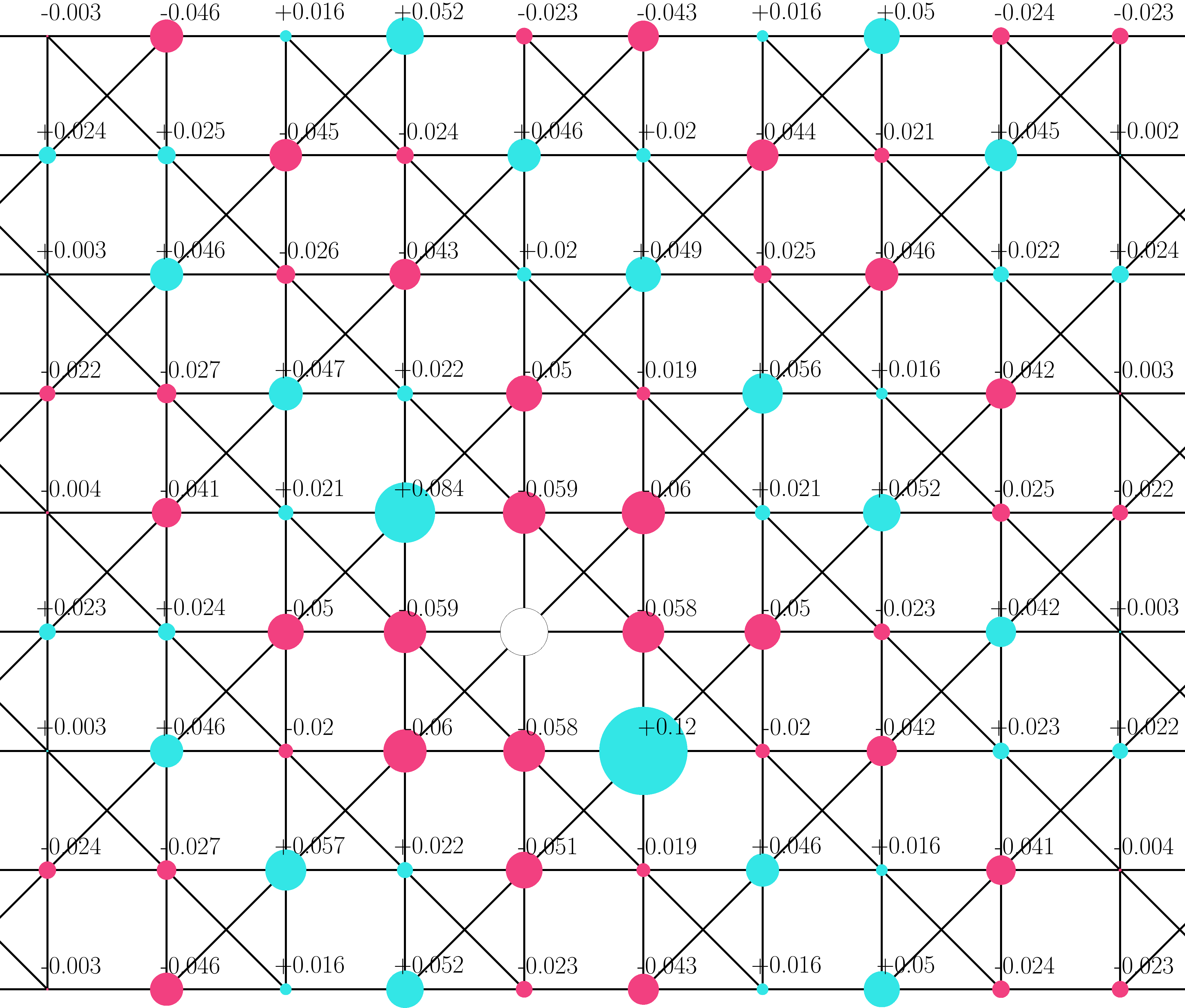}
\caption{(Color online) Connected $S^z$ correlations between the reference site (black circle) and the neighboring ones computed using DMRG at $m=1/2$ on a $16\times 8$ cylinder using CBC. Top/bottom panels correspond respectively to $\Delta=1$ and $5$. Only data within the bulk are shown. Positive and negative values are shown respectively with blue and red colors.}
\label{fig:XXZ}
\end{figure}

By computing the ground-state with $m=1/2$ at $\Delta=5$ with DMRG, we have observed  that there are qualitatively distinct features with respect to the SU(2) case ($\Delta=1$): both local magnetization and bond strengths become uniform in the bulk (data not shown), i.e. there is no evidence for the simple VBC that was found at $\Delta=1$. To clarify the different nature in these ground-states, we plot in Fig.~\ref{fig:XXZ} the connected correlations $\langle S^z_i S^z_j\rangle - \langle S^z_i \rangle \langle S^z_j\rangle$ in both regimes. 
For $\Delta=1$, we have previously shown that the ground-state is a VBC with inhomogeneous sites, see Fig.~\ref{fig:dmrg2} so that in principle we would need to compute different correlations depending on the reference site.
 However, this is for illustration only since we already know the nature of this VBC and these correlation data simply reflect that we have an almost product state so that correlations between plaquettes are very small, and inside one plaquette, they show the expected pattern for (here) a singlet state. On the contrary, for $\Delta=5$, Fig.~\ref{fig:XXZ} provides a strong evidence that the ground-state at $m=1/2$ is similar to the columnar phase of the effective QDM: in the spin language, we have diagonal lines that repeat a simple $\downarrow \uparrow \uparrow \uparrow$ pattern, thus explaining the observed pattern. So in the spin language, it does correspond to a ferrimagnetic phase, not a VBC one.
 
\section{Conclusion}\label{sec:conclusion}
We have provided strong numerical evidence in favor of the existence of magnetization plateaux on the spin-1/2 Heisenberg model on the 2d checkerbord lattice for $m=0$, $1/4$, $1/2$ and $3/4$ of its saturation value. While the $m=0$ plateau (due to the finite spin gap) was previously known from the literature~\cite{Palmer2001,Canals2002,Fouet2003,Brenig2002,Tchernyshyov2003,Berg2003} and corresponds to a 2-fold degenerate VBC, we find that the three others are well described by a 4-fold degenerate VBC, analogous to the exact localized magnon eigenstate that can be constructed at $m=3/4$.~\cite{Richter2004} Thus, the situation is rather similar to another famous corner-sharing geometry, namely the kagom\'e lattice, where the same phenomenology has been recently observed.~\cite{Capponi2013,Nishimoto2013} It seems that the finite-field situation can be better understood from the large field limit, which is more amenable to theoretical techniques presumably, or less frustrated in a sense. Moreover, these product states can also be interpreted as having quantized spin imbalance (obtained by measuring magnetization on different blocks), which could be a interpreted as a remnant of the classical degeneracy through an order-by-disorder mechanism.~\cite{Plat2015}

We would like also to comment about the adiabatic connections (or not) between these plateaux phases and similar ones that have been observed in different context.~\footnote{This was for instance recently investigated for the kagome lattice~\cite{Huerga2016,Kshetrimayum2016}} For $m=0$, there is a recent numerical evidence of plaquette phase persisting  in the antiferromagnetic XY limit.~\cite{Laeuchli2015} 
In the opposite XXZ limit with dominant Ising interaction, the $m=0$ and $m=1/2$ low-energy configurations can be easily seen to be in a one-to-one correspondence respectively with quantum loop~\cite{Shannon2004} and quantum dimer configurations on a square lattice. Both effective constrained models are of the Rokhsar-Kivelson type~\cite{Rokhsar1988} with purely kinetic terms, and have been investigated quite extensively. This quantum loop model (also known as square ice) has a 2-fold degenerate plaquette ground state~\cite{Shannon2004,Banerjee2013}, i.e. a similar structure as our $(0,0)$ product state, hence pointing to a robust feature present for any anisotropy in the XXZ sense. On the contrary, the quantum dimer model with purely kinetic term has a 4-fold degenerate columnar ground-state~\footnote{Let us emphasize that the complete phase diagram of the quantum dimer model on the square lattice is still an ongoing investigation, see Refs.~\onlinecite{Syljuasen2006,Banerjee2014,Schwandt,Banerjee2016} and references therein.}, which, when translated into spin language, would correspond to a ferrimagnetic state with fixed local magnetization $\pm 1/2$, i.e. qualitatively different from the resonating state that we have found in the SU(2) case. Therefore, we predict the existence of a quantum phase transition when increasing the anisotropy of the XXZ model between a VBC and an ordered ferrimagnetic phase, which is confirmed numerically see Sec.~\ref{sec:qpt}. The nature of this phase transition is potentially interesting (continuous vs first order, see for instance Ref.~\onlinecite{DEmidio2016}) but a complete analysis is postponed to a future study. 
In a similar manner, it would be natural to investigate the fate of $m=1/4$ and $3/4$ plateaux when moving away from the SU(2) case. In addition, it would be interesting to prove whether supersolidity can be stabilized in the vicinity of some of these plateaux, a phenomenon which is common on frustrated lattices but not present on the checkerboard lattice for interacting hardcore bosons with non-frustrated hopping.~\cite{Wessel2012}

As far as fermionic Hubbard-like models on the same lattice are concerned,  it has been shown that various VBC can also be stabilized at commensurate fillings~\cite{Indergand2006,Indergand2007,Poilblanc2007}. Generalization of spin models with SU($N$) symmetry (using fundamental representation at each site) can also host other kinds of valence bond solid states.~\cite{Corboz2012} Hence, a global picture of the effect of strong quantum correlations is emerging thanks to all of these studies and provides new exotic phases compared to the classical ones.

\acknowledgments{
This work was performed using HPC resources from GENCI (Grant No. x2015050225 and No. x2016050225) and CALMIP. The author thanks Institut Universitaire de France (IUF) for funding when this work started and the Condensed Matter Theory Visitors Program at Boston University for financial support during the completion of this manuscript. Discussions are acknowledged with Andreas L\"auchli and Xavier Plat.}

{\it Added note:}  During completion of this work, a related DMRG study has appeared~\cite{Morita2016} and fully agrees on the nature and stability of the proposed VBCs here.

\end{document}